\documentclass[aps,pra,twocolumn,superscriptaddress,showpacs,floatfix]{revtex4-1}
\usepackage{graphicx,dcolumn,bm,hyperref,amsmath,amssymb,xspace,epsfig,float,array,multirow,amsfonts,color}
\usepackage[caption=false]{subfig}
\usepackage[vcentermath]{youngtab}
\usepackage{braket}
\usepackage[normalem]{ulem}
\usepackage{ytableau}

\begin{document}

\title{Graph-theory treatment of one-dimensional strongly repulsive fermions}

\author{Jean Decamp}
\affiliation{MajuLab, CNRS-UCA-SU-NUS-NTU International Joint Research Unit, Singapore}
\affiliation{Centre for Quantum Technologies, National University of Singapore, 117543 Singapore, Singapore}
\affiliation{Department of Physics, National University of Singapore, 2 Science Drive 3, Singapore 117542, Singapore}

\author{Jiangbin Gong}
\affiliation{Department of Physics, National University of Singapore, 2 Science Drive 3, Singapore 117542, Singapore}

\author{Huanqian Loh}
\affiliation{Centre for Quantum Technologies, National University of Singapore, 117543 Singapore, Singapore}
\affiliation{Department of Physics, National University of Singapore, 2 Science Drive 3, Singapore 117542, Singapore}

\author{Christian Miniatura}
\affiliation{MajuLab, CNRS-UCA-SU-NUS-NTU International Joint Research Unit, Singapore}
\affiliation{Centre for Quantum Technologies, National University of Singapore, 117543 Singapore, Singapore}
\affiliation{Department of Physics, National University of Singapore, 2 Science Drive 3, Singapore 117542, Singapore}
\affiliation{School of Physical and Mathematical Sciences, Nanyang Technological University, 637371 Singapore, Singapore}
\affiliation{Yale-NUS College, 16 College Avenue West, Singapore 138527, Singapore}
\affiliation{Universit\'{e} C\^{o}te d'Azur, CNRS, INPHYNI, Nice, France}

\date{\today}

\begin{abstract}
One-dimensional atomic mixtures of fermions can effectively realize spin chains and thus constitute a clean and controllable platform to study quantum magnetism. Such strongly correlated quantum systems are also of sustained interest to quantum simulation and quantum computation due to their computational complexity.  In this article, we exploit spectral graph theory to completely characterize the symmetry properties of one-dimensional fermionic mixtures in the strong interaction limit.  We also develop a powerful method to obtain the so-called Tan contacts associated with certain symmetry classes.  In particular, compared to brute force diagonalization that is already virtually impossible for a  moderate number of fermions, our analysis enables us to make unprecedented efficient predictions about the energy gap of complex spin mixtures.  Our theoretical results are not only of direct experimental interest, but also provide important guidance for the design of adiabatic control protocols in strongly correlated fermion mixtures.
\end{abstract}

\maketitle

%%%%%%%%%%%%%%%%%%%%%%%%%%%%%%%%%%%%%%%%%%%%%%%%%%%%%%%%%%%%%%%%%%%%%%

From quantum magnetism to the highly debated high-temperature superconductivity, many spectacular phenomena observed in condensed matter physics emerge from strong interactions among particles with a spin degree of freedom \cite{assa}. When constrained to one dimension, the effect of correlations is even higher, leading to counter-intuitive behaviour such as spin-charge separation and the so-called fermionization \cite{giamarchi}. Although the study of such strongly correlated systems is a notoriously complex task, recent experimental realizations of one-dimensional (1D) systems involving ultracold atomic fermions with $\kappa\ge 2$ spin degrees of freedom offer exciting opportunities, both in terms of our fundamental understanding of 1D quantum magnetism and in the prospect of quantum technological applications \cite{Gorshkov2010,Ye2014,Foelling2014,Pagano2014,Loft2016,Marchukov2016}. Indeed, in the limit of strong repulsion, it has been shown that these systems are equivalent to spin chains whose interaction parameters are experimentally tunable with the external potential \cite{Volosniev2014,Deuretzbacher2014,Murmann2015}. In particular, this non-trivial observation provides a mapping to a matrix diagonalization problem. However, the algebraic structure of this matrix has never been clearly identified. Therefore, although many recent theoretical articles have studied this model, the results were typically limited to 6 particles, beyond which the complexity of the system seemed to be an impassable barrier \cite{Cui2013,Massignan2015,Grining2015,Beverland2016,Decamp2016a,Decamp2016b,Pecak2017,Sowiski2019}.

In this article, we claim that the spin chain model associated with 1D strongly repulsive fermionic mixtures has a natural interpretation in terms of spectral graph theory. This hitherto unobserved connection to a well-studied mathematical branch \cite{Mohar1991,Bacher1994,Friedman2000,Cesi2010,Poignard2018} provides a general and rigorous framework, which enables us to completely elucidate the symmetry structure of the spectrum for arbitrary external potentials and numbers of particles. This has strong implications: for example, it allows us to prove a generalized form of the Lieb-Mattis theorem, which implies in particular that the ground-state of the system is unmagnetized \cite{LiebMattisPR}. Once again, this theorem has previously only been conjectured for 6-particles mixtures confined in a harmonic potential \cite{Decamp2016a,Decamp2016b}. More importantly, this framework and the symmetry structure we deduced allows one to split the problem, which is extremely complex, into lower dimensional irreducible representations. This enables us in particular to compute the energy gap with polynomial efficiency instead of exponential, thus providing a simple answer to a critical outstanding problem for adiabatic quantum computing \cite{Zhang2014,Albash2018}.

\begin{figure}
	\subfloat{\label{ex4a}\includegraphics[width=1\linewidth]{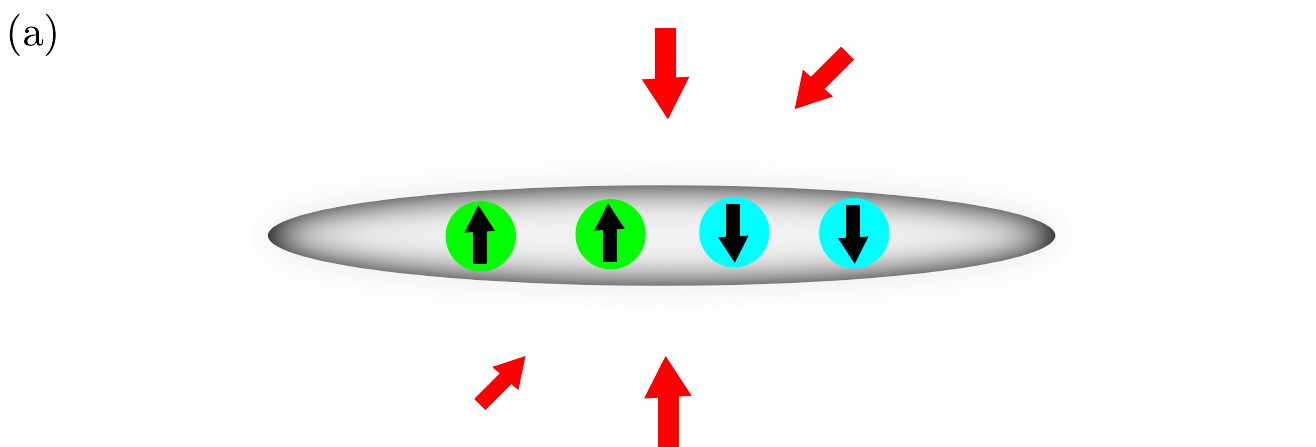}}\\
	\subfloat{\label{ex4b}\includegraphics[width=1\linewidth]{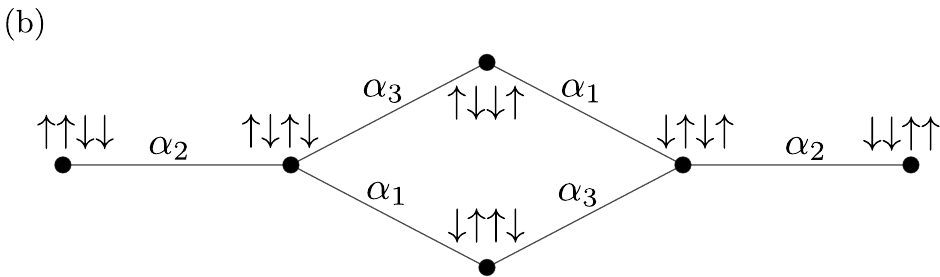}}
	\caption{\label{ex4}(color online). A spin mixture with two spin up and two spin down fermions. (a) Typical experimental realization, where the fermionic atoms are confined to 1D by interfering optical lattice beams along two dimensions. (b) Graph theory interpretation of the system. The vertices are labelled by the permutations of the mixture, or snippets, and an edge by the nearest-neighbour exchange constant $\alpha_k$ (c.f. Eq.~\eqref{alphavolo}) if it connects two vertices that are equal up to a transposition in positions $(k,k+1)$. The $V^{(2,2)}$ matrix defined in Eq.~\eqref{matrixvolo} of the main text  is then the Laplacian matrix of $X(\mathfrak{S}_{(2,2)}\subset \mathfrak{S}_4,S_c)$.}
\end{figure}

\section{Model and quantities of interest}

We consider a 1D system of $N$ particles divided in $\kappa$ fermionic components (e.g. spin orientations) with populations given by the partition $\nu\equiv(N_1,\ldots,N_{\kappa})$ of $N$, i.e. such that $N_1\ge\ldots\ge N_{\kappa}>0$ and $N_1+\cdots+ N_{\kappa}=N$. We suppose that it is  $SU(\kappa)$-symmetric, so that all the particles have the same mass $m$, interact via the same $\delta$-type potential with interaction strength $g_{\mathrm{1d}}$, and are submitted to the same confining potential $V_{\mathrm{ext}}(x)$, which can be arbitrary. It can be described by the following Hamiltonian:
\begin{equation}
\label{maineq}
\hat{H}=\sum_{j=1}^N\left(-\frac{\hbar^2}{2m} \frac{\partial^2}{\partial x_j^2}+V_{\mathrm{ext}}(x_j)\right)+g_{\mathrm{1d}}\sum_{i<j}\delta(x_i-x_j)
\end{equation}
where $x_j$ is the coordinate of particle $j$ and we suppose that the $N_1$ first particles belong to component $1$, etc (Fig.~\ref{ex4a}). Such a model has been realized experimentally using a harmonic potential to confine ${}^{173}\mathrm{Yb}$ ultracold atoms, whose ground-state's purely nuclear spin ($I=5/2$) guarantees the absence of spin flip collisions and $SU(\kappa)$ symmetry with $\kappa\in\{2,\ldots,6\}$ \cite{Pagano2014}. In the following, we will consider the strongly repulsive limit $g_{\mathrm{1d}}\to+\infty$, which can be achieved by means of a confinement-induced resonance \cite{Olshanii1998}. In the formal $1/g_{\mathrm{1d}}=0$ regime, the system is said to be fermionized: it has the same spectrum (up to the degeneracy) as the spinless $N$-particle fermionic system, with a totally antisymmetric Slater determinant $\psi_A$ of one-particle orbitals as a wavefunction. However, its permutation symmetry is \textit{a priori} different. Following \cite{Volosniev2014,Deuretzbacher2014} and generalizing Girardeau's  Bose-Fermi mapping \cite{Girardeau1960}, we write the wavefunction $\psi$ of the system in the Bethe ansatz-like form
\begin{equation}
\label{Psivolo}
\psi=\sum_{P\in \mathfrak S_N}a_P~\theta(x_{P1}<\cdots<  x_{PN})~\psi_A(x_1,\ldots,x_N),
\end{equation}
 where $\mathfrak S_N$ is  the permutation group of $\{1,\ldots,N\}$, $P(1,\ldots,N)=(P1,\ldots,PN)$, $\theta(x_1<\cdots<  x_N)$ is equal to 1 if $x_1<\cdots<  x_N$ and 0 otherwise, and the vector $(a_P)\in\mathbb{R}^{N!}$ totally encodes the permutation symmetry of the state. For a given mixture $\nu=(N_1,\ldots,N_{\kappa})$ the Pauli principle reduces the number of different coefficients to  $D_{\nu}=N!/(N_1!N_2!\cdots N_{\kappa}!)$.
The $D_{\nu}$-dimensional vector space of classes of sectors that are equivalent up to a permutation of identical particles is  called the snippet space  \cite{Fang2011}. Noting that the system is no longer degenerate in the vicinity of the $1/g_{\mathrm{1d}}=0$ point, one can write a first-order expansion of the energy in $1/g_{\mathrm{1d}}$:
\begin{equation}
\label{perten}
E(1/g_{\mathrm{1d}})=E_A-K^{\nu}\left[(a_P)\right]/g_{\mathrm{1d}}+o(1/g_{\mathrm{1d}}),
\end{equation}
where $E_A$ is an eigenenergy of the non-interacting Hamiltonian associated with $\psi_A$ and the energy slope $K^{\nu}\left[(a_P)\right]=-\lim_{g_{\mathrm{1d}}\to\infty}\frac{\partial E}{\partial g_{\mathrm{1d}}^{-1}}=g_{\mathrm{1d}}^2\frac{\partial E}{\partial g_{\mathrm{1d}}}$ is a functional of $(a_P)$.  This quantity is equivalent to the so-called Tan's contact, a pivotal and experimentally accessible quantity in $\delta$-interacting systems that governs the large $p$ behaviour of the momentum distribution $n(p)$ \cite{Olshanii2003,Tan2008a,Hui11,Zwerger2011,Stewart2010,Chang2016}. Then, it can be shown that the values of $(a_P)$ and $K^{\nu}\left[(a_P)\right]$ corresponding to the states belonging to the same degenerate manifold can be found by solving the following eigenvalue equation \cite{Volosniev2014,notespinchain}:
\begin{equation}
\label{diagkeq}
V^\nu\vec{a}=K^{\nu}\vec{a},
\end{equation}
where $\vec{a}$ is the vector of the $D_{\nu}$ independent $a_P$ coefficients and $V^\nu$ is a $D_{\nu}\times D_{\nu}$ real symmetric matrix defined in the snippet space by
\begin{equation}
\label{matrixvolo}
V^\nu_{PQ}=\left\{
\begin{array}{ll}
-\alpha_{P,Q} & \mbox{if } P\ne Q \\ \sum_{R\ne P}\alpha_{P,R}& \mbox{if  } P=Q
\end{array}
\right..
\end{equation}
The so-called nearest-neighbour exchange constants $\alpha_{P,Q}$ are given by
\begin{equation}
\label{alphavolo}
\alpha_k=\int_{x_1<\cdots<x_N} dx_1\ldots dx_N~\delta(x_k-x_{k+1})\left|\frac{\partial\psi_A}{\partial x_k}\right|^2
\end{equation}
if $P$ and $Q$ are equal up to a transposition in positions $k$ and $k+1$ with the particles in $k$ and $k+1$ belonging to different components, and $\alpha_{P,Q}=0$ otherwise. Their values depend exclusively on $V_{\mathrm{ext}}(x)$ and $N$, which makes them experimentally tunable. Then, the ground state  corresponds to the vector $\vec{a}_{max}$ associated with the largest eigenvalue $K^{\nu}_{max}$ of $V^\nu$, and the other states are given by the eigen-decomposition of $V^\nu$. 

\section{Connection with spectral graph theory}

In what follows, we explain how symmetry and spectral properties of the spin chain model can be elucidated using spectral graph theory. The basic notions of finite group and representation theories are available from references \cite{Fulton2004,Liebeck,Hamermesh_book}.

Given a finite group $G$ and a generating subset $S$ of $G$, it is possible to associate a graph $X(G,S)$ called a Cayley graph, where each vertex of $X(G,S)$ is labelled by the elements of $G$ and such that two vertices $(g,g')$ are connected by an edge if and only if there is $s\in S$ such that $g'=sg$. If we assign a weight $w_s>0$ to each $s\in S$, the graph $X(G,S)$ is weighted \cite{note1a}. If we also consider a subgroup $H$ of $G$, one can associate with $G$, $H$ and $S$ a so-called Schreier graph $X(H\subset G,S)$, whose vertices are indexed by the left cosets $gH$ and with the edges given by $(gH,sgH)$ where $g\in G$, $s\in S$ and $gH\ne sgH$ \cite{note2}. These graphs are an essential tool in combinatorial and geometric group theory \cite{Brouwer2012}. Then, our connection can be established as follows: given a fermionic mixture defined by a partition $\nu\equiv(N_1,\ldots,N_{\kappa})$ of $N$, the $V^\nu$ matrix defined in Eq.~\eqref{matrixvolo} is equal to the Laplacian matrix of the weighted Schreier graph $X(\mathfrak{S}_{\nu}\subset \mathfrak{S}_N,S_C)$. Here, $\mathfrak{S}_{\nu}=\mathfrak{S}_{N_1}^{\nu}\times\cdots\times\mathfrak{S}_{N_{\kappa}}^{\nu}$ is the Young subgroup associated with $\nu$, with $\mathfrak{S}_{N_i}^{\nu}$ the set of permutations $P\in\mathfrak{S}_N$ such that $P(i)=i$ if $i$ does not belong to $\{N_1+\cdots +N_{i-1}+1,\dots,N_1+\cdots + N_{i-1}+N_i\}$ \cite{noteys}; and  $S_C\equiv\{(1,2),(2,3),\ldots,(N-1,N)\}$ is the set of nearest-neighbour transpositions, where each $(k,k+1)\in S_C$ is associated with a weight $\alpha_k$ as defined in Eq.~\eqref{alphavolo}. An illustration in the case of a mixture of two spin up and two spin down is given in Fig.~\ref{ex4b}, and all the non-trivial graphs  $X(\mathfrak{S}_{\nu}\subset \mathfrak{S}_5,S_c)$ in the case of $N=5$ particles are displayed in Fig.~\ref{figgraphs}. Interestingly, this correspondence provides an interpretation of the Laplacian matrix $V^\nu$ as the generator of a random walk known as an interchange process \cite{Cesi2010,Lovász93randomwalks}: The particles can be seen as a deck of $N$ cards, and, at rate 1, two adjacent cards $(k,k+1)$ are selected with a probability given by $\alpha_k$ and exchanged.

\section{General structure of the spectrum}

\begin{figure}
	\subfloat[\label{path}]{\includegraphics[width=0.4\linewidth]{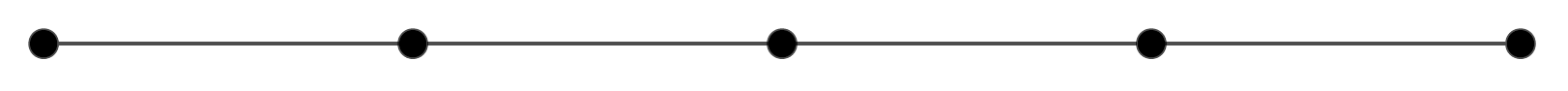}}\hfill
	\subfloat[\label{b}]{\includegraphics[width=0.4\linewidth]{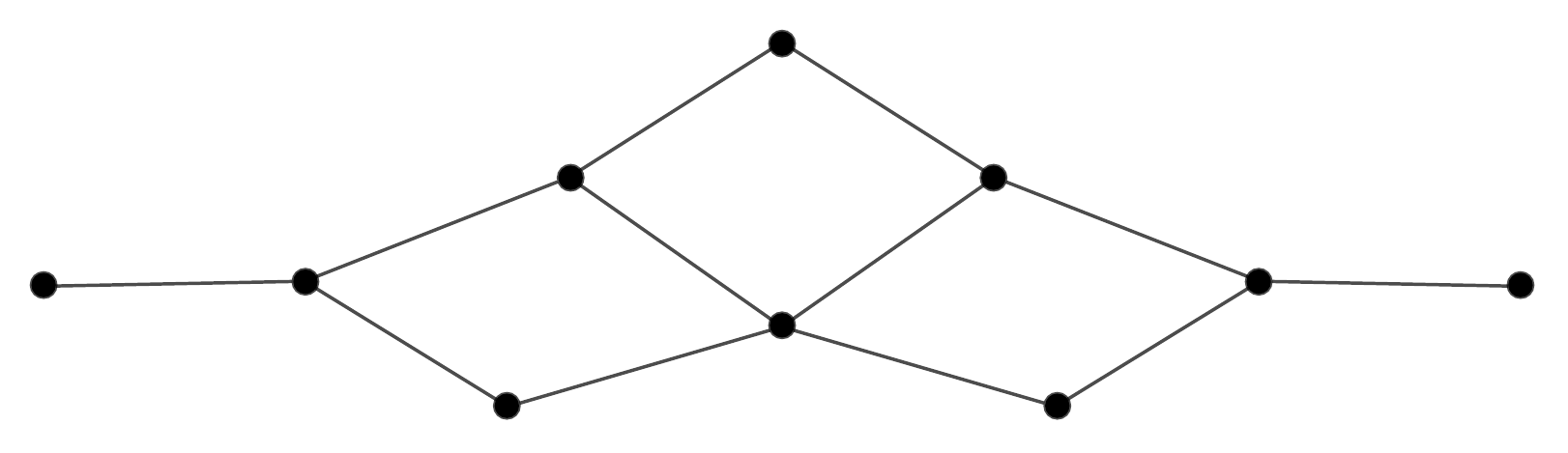}}\\
	\subfloat[\label{311}]{\includegraphics[width=0.4\linewidth]{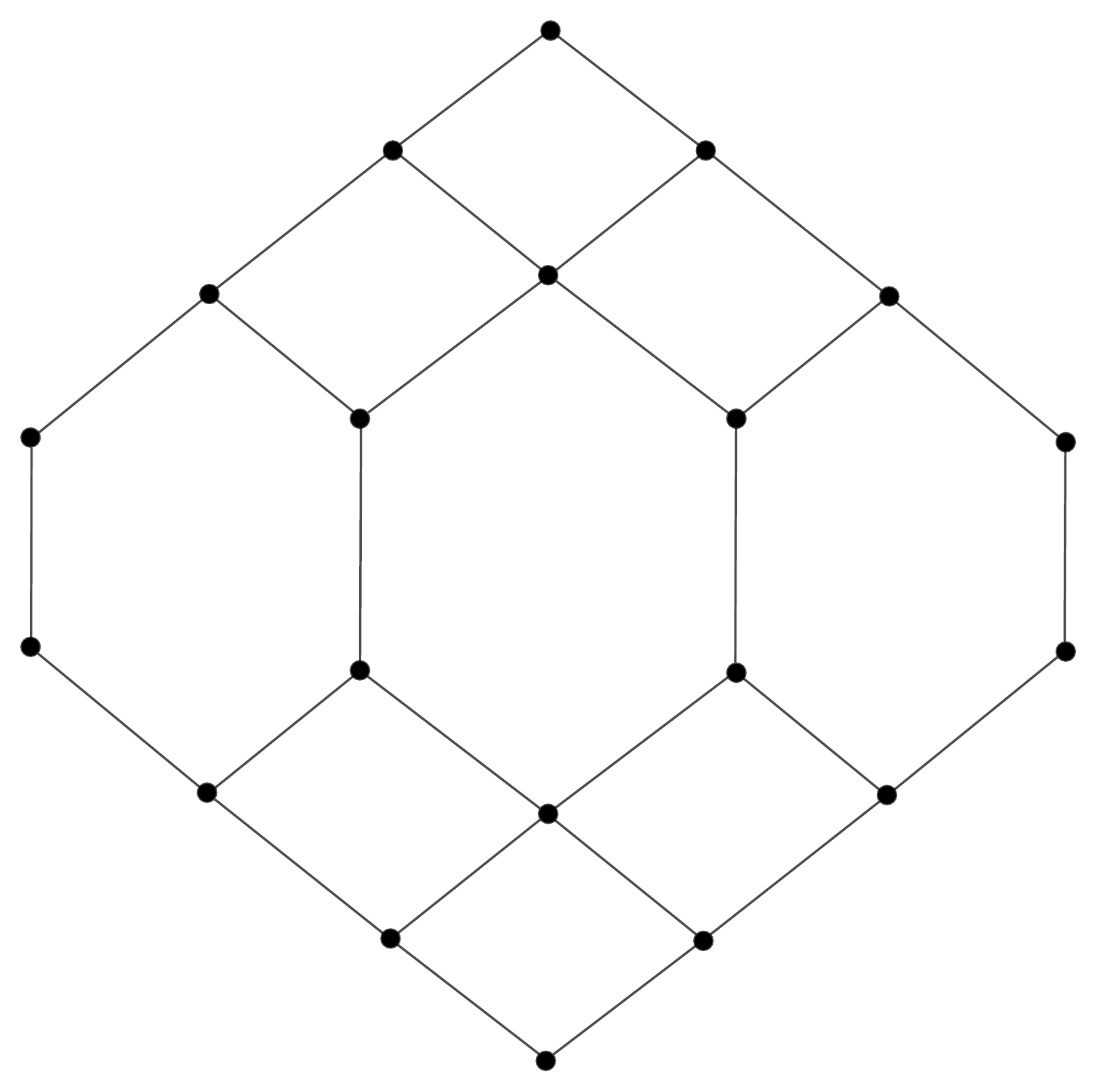}}\hfill
	\subfloat[\label{221}]{\includegraphics[width=0.4\linewidth]{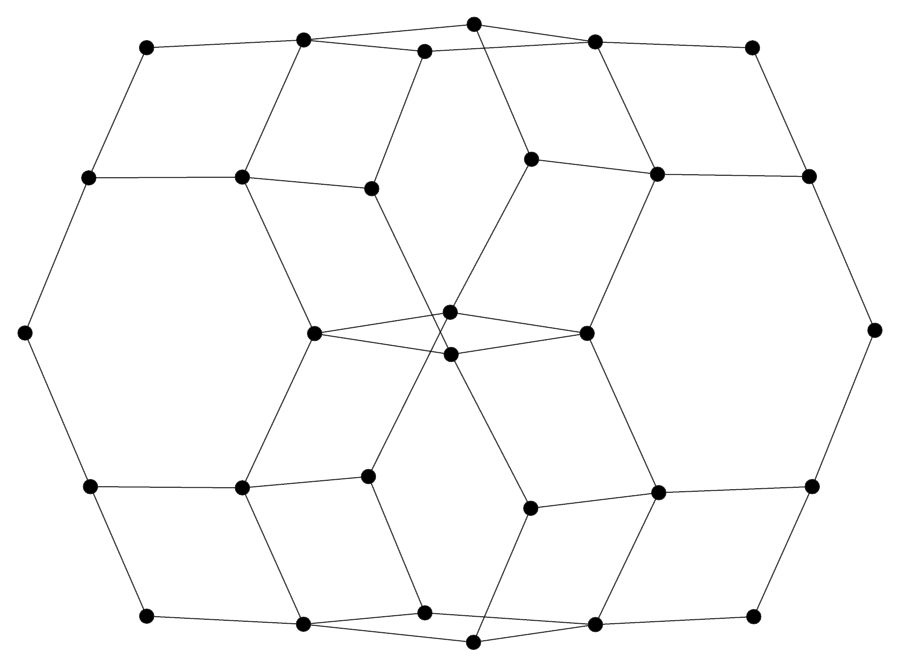}}\\
	\subfloat[\label{2111}]{\includegraphics[width=0.4\linewidth]{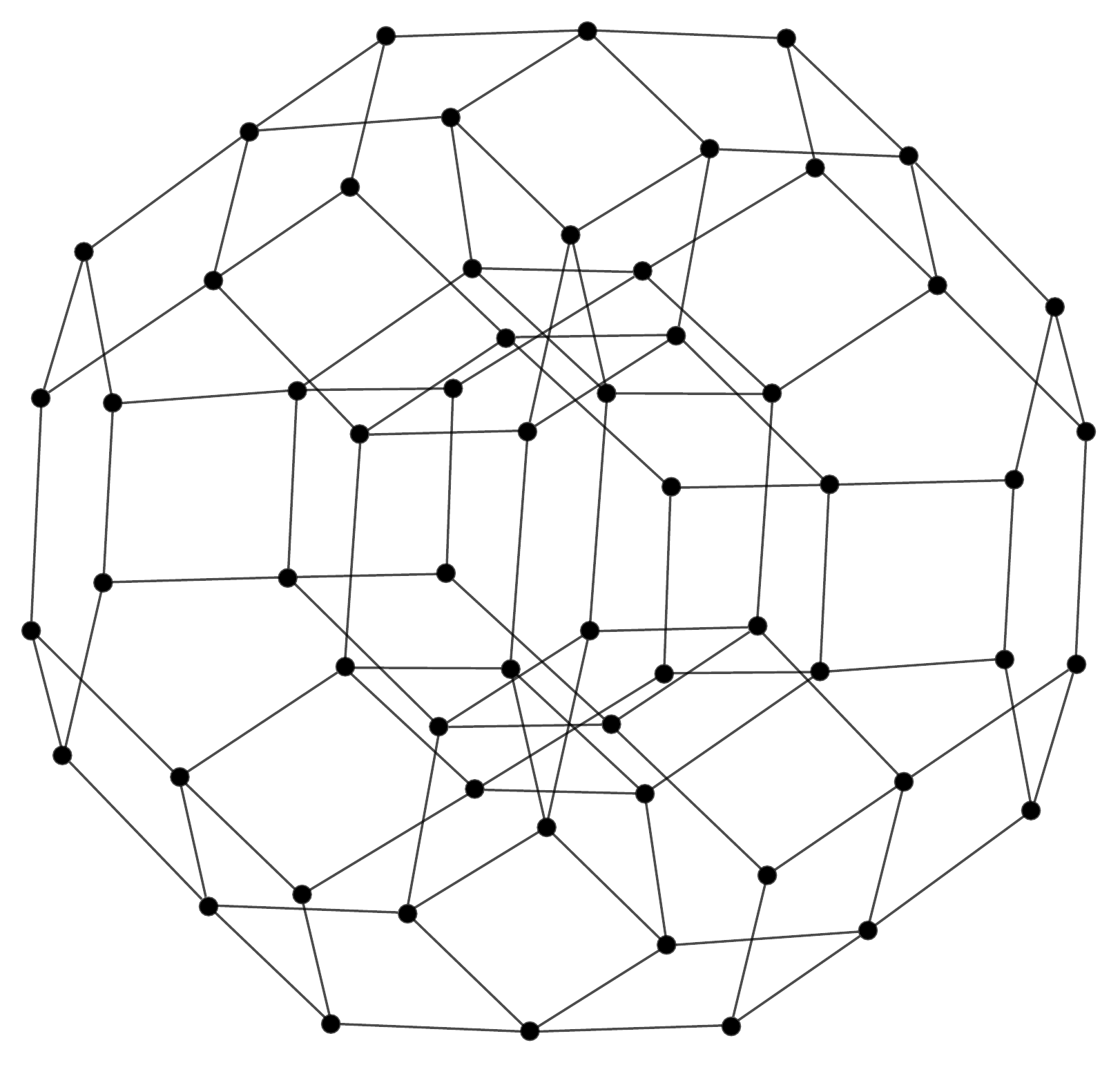}}\hfill
	\subfloat[\label{11111}\label{phedron}]{\includegraphics[width=0.4\linewidth]{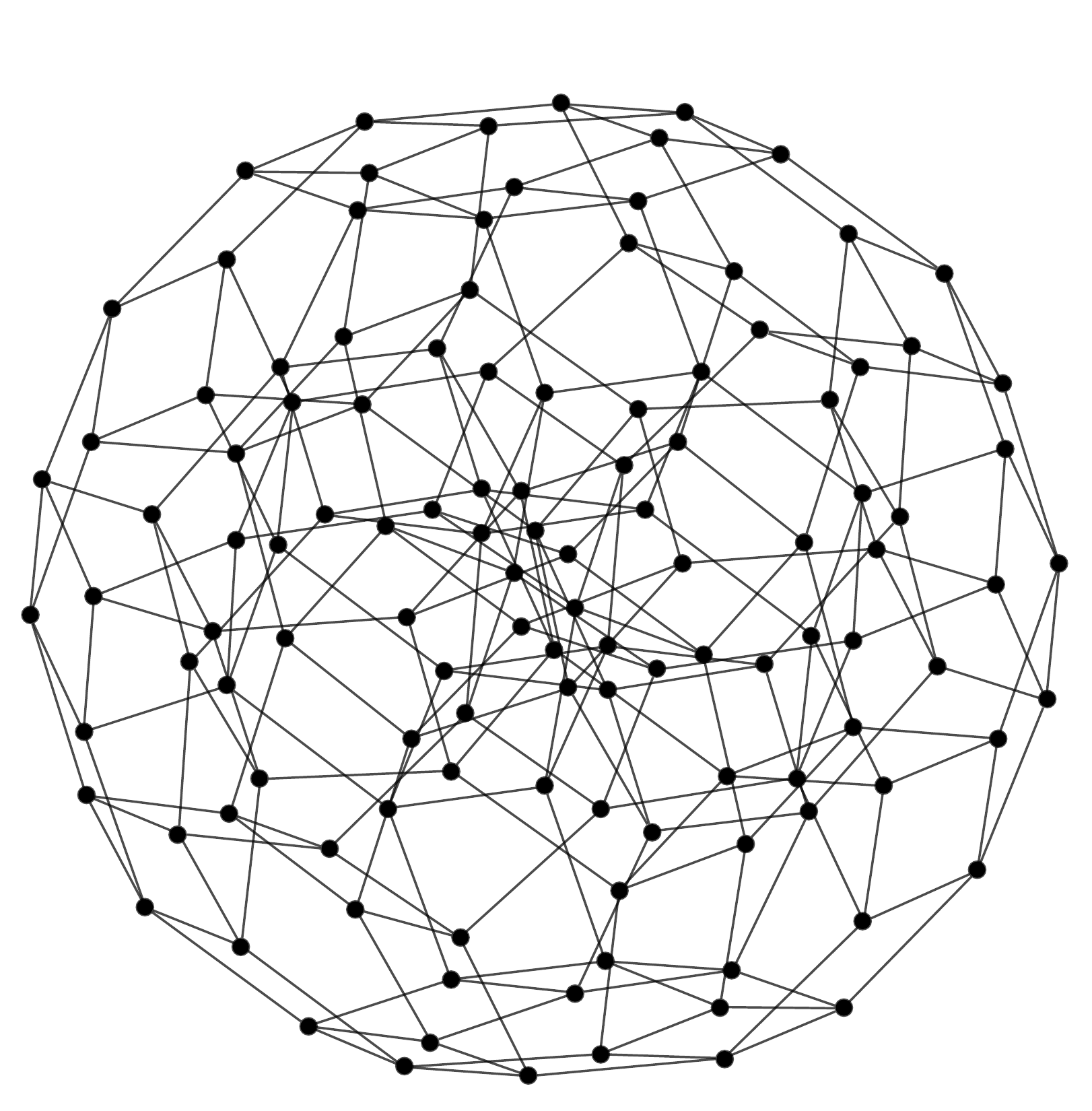}}
	\caption{\label{figgraphs}Panels (a) to (f): Schreier graphs $X(\mathfrak{S}_{\nu}\subset \mathfrak{S}_5,S_c)$ corresponding to the partitions $\nu=$ $(4,1)$, $(3,2)$,  $(3,1,1)$, $(2,2,1)$, $(2,1,1,1)$, $(1,1,1,1,1)$ of $N=5$ (resp.). The one-component case $\nu=(5)$ is trivial. In the case (f) where there are as many components as particles, the corresponding Cayley graph $X(\mathfrak{S}_5,S_c)$ is called a permutohedron \cite{Bowman1972}. As argued in the main text, the spectral gap in this case is the same as in the much simpler case (a) of the  path graph.}
\end{figure}

Considering two Schreier graphs $X(H\subset G, S)$ and $X(K\subset G, S)$ such that $K\subset H$, one can show that the latter is a covering graph of the former, and therefore the Laplacian spectrum $\sigma\left(X(H\subset G,S)\right)$ is contained in $\sigma\left(X(K\subset G,S)\right)$ \cite{Bacher1994,Friedman2000}. In our situation, for example, the fact that $\mathfrak{S}_{(2,1,1,1)}\cong\mathfrak{S}_{2}\subset\mathfrak{S}_{3}\cong\mathfrak{S}_{(3,1,1)}$ implies that the graph in  Fig.~\ref{2111} is a covering of the one in Fig.~\ref{311}, and that $\sigma\left( V^{(3,1,1)}\right)\subset\sigma\left( V^{(2,1,1,1)}\right)$. More generally, at fixed $N$, $\sigma\left( V^{\nu}\right)\subset\sigma\left( V^{(1,1,\ldots ,1)}\right)$ for every partition $\nu$ of $N$. In physical terms, this means that the high-spin limit $N=\kappa$ where there are as many particles as components encompasses all the other cases of mixtures. The resulting question is to determine, for a given mixture $\nu$, to which subset of $\sigma\left( V^{(1,1,\ldots ,1)}\right)$ the spectrum of $V^{\nu}$ corresponds. To answer this question, one has to note that the $V^{\nu}$ matrices can be seen as representations of the symmetric group $\mathfrak{S}_N$. In the case where $\nu=(1,1,\ldots ,1)$, each vertex of the Cayley graph $X(\mathfrak{S}_N,S_c)$ corresponds uniquely to a permutation of the mixture. Then, $V^{(1,1,\ldots ,1)}$ acts on the vector space whose basis is indexed by the elements of $\mathfrak{S}_N$, known as the group algebra. Thus, this matrix can be expressed in terms of the so-called regular representation $\rho$, which is equivalent to the direct sum of all the irreducible representations (irreps) $\rho_{\mu}$ labelled by the partitions  $\mu$ of $N$ \cite{James1984}. Denoting by $e_k$ the element of the group algebra corresponding to a transposition $(k,k+1)$ and by $\mathrm{IR}$ the set of all the irreps of $\mathfrak{S}_N$, we get that $\sigma\left( V^{(1,1,\ldots ,1)}\right)$ is equal to
\begin{equation}
\label{spectrum}
	\bigcup_{\rho_\mu\in\mathrm{IR}}\left\{d-\lambda~:~\lambda\in\sigma\left(\rho_\mu\left(\sum_{k=1}^{N-1}\alpha_ke_k\right)\right)\right\},
\end{equation}
where  $d=\sum_{k=1}^{N-1}\alpha_k$. Then, in the case of an arbitrary mixture $\nu$, the key observation is that each snippet can  be uniquely associated with a so-called tabloid of shape $\nu$, and therefore that $V^{\nu}$ is acting on the vector space indexed by the tabloids, known as the permutation module $M^{\nu}$ \cite{note1sm}. According to the so-called Young's rule \cite{James1984}, $\sigma\left(V^{\nu}\right)$ is thus given by Eq.~\eqref{spectrum}, where $\rho_{\mu}$ is now taken over the elements of $\mathrm{IR}$ such that  $\mu\trianglerighteq\nu$. Here, $\trianglerighteq$ is the dominance order: $\left[\mu_1,\ldots,\mu_r\right]\trianglerighteq\left[\nu_1,\ldots,\nu_r\right]$ if $\mu_1+\cdots+\mu_k\ge\nu_1+\cdots+\nu_k$ for all $k$. This central result, which can be understood as a consequence of the Pauli principle, completely elucidates the symmetry structure of the spectrum.

\section{Symmetry ordering}

Denote by $K_{max}^{[\mu]}$ the maximum eigenvalue of $V^{(1,1,\ldots ,1)}$ belonging to the symmetry class $[\mu]$. From our previous discussion, we see that $K_{max}^{[\mu]}\in\sigma\left(V^{\mu}\right)$ (since $\mu\trianglerighteq\mu$) and thus that $K_{max}^{[\mu]}\le K_{max}^{\mu}$. In fact, using the fact that the graph $X(\mathfrak{S}_{\nu}\subset \mathfrak{S}_N,S_C)$ is bipartite, one can always construct an eigenvector with eigenvalue $K_{max}^{\mu}$ that belongs to the symmetry class $[\mu]$, so that $K_{max}^{[\mu]} = K_{max}^{\mu}$ \cite{note2sm}. Therefore, we have shown that
\begin{equation}
\label{klieb}
\mu\trianglerighteq\mu'\quad\implies\quad K_{max}^{[\mu]}\le K_{max}^{[\mu']}.
\end{equation}
This fact has been numerically \cite{Decamp2016a,Decamp2016b} and experimentally \cite{Pagano2014} observed. It can be interpreted as a generalized Lieb-Mattis theorem \cite{LiebMattisPR}, c.f. Eq.~\eqref{perten}, which is salient in the context of quantum magnetism. Intuitively, the ground state is as spatially symmetric as possible and its total spin is minimized.

\section{Peculiar eigenvalues and spectral gap}

From basic properties of the Laplacian matrix, we see that $0\in\sigma\left(V^{\nu}\right)$ for every mixture $\nu$. Moreover, the fact that $X(\mathfrak{S}_N,S_C)$ is bipartite and $d$-regular implies that $K_{max}^{(1,1,\ldots,1)}=2d$, and more generally that $K\in\sigma\left(V^{(1,1,\ldots,1)}\right)$ if and only if $2d-K\in\sigma\left(V^{(1,1,\ldots,1)}\right)$ \cite{Brouwer2012}. The simple eigenvalues $0$ and $2d$ are associated with the trivial $\left[N\right]$ and sign $\left[1,1,\ldots,1\right]$ irreps, respectively.

In \cite{Bacher1994}, Bacher has studied $\sigma\left(X(\mathfrak{S}_N,S_C)\right)$ in the unweighted case where $\alpha_k=1$ for all $k$, obtaining in particular an expression for the spectral gap, which is deeply related to the geometrical properties of $X(\mathfrak{S}_N,S_C)$ \cite{Chung1996,Godsil2001}. His argument is based on a mapping between this graph and the cartesian product of $r$ path graphs $X(\mathfrak{S}_{(N-1,1)}\subset\mathfrak{S}_N,S_C)$ (c.f. Fig.~\ref{path}). 

We now proceed to generalize this non-trivial result to the weighted case.
 Let $\sigma\left(V^{(N-1,N)}\right)=\left\{0<\lambda_2<\lambda_3<\cdots<\lambda_N\right\}$, where $V^{(N-1,N)}$ is given by the Laplacian matrix of the weighted path graph:
\small
\begin{equation}
\label{matpath}
\begin{pmatrix}
\alpha_1 & -\alpha_1  &  & & \\
-\alpha_1& \alpha_1+\alpha_2 & -\alpha_2  & & \\
 & \ddots & \ddots & \ddots & \\
  & & -\alpha_{N-2} & \alpha_{N-1}+\alpha_{N-2} & -\alpha_{N-1}\\
& &  & -\alpha_{N-1} & \alpha_{N-1}
\end{pmatrix}.
\end{equation}
\normalsize
Then, $\sigma\left(V^{(1,1,\ldots,1)}\right)$ contains the following eigenvalue
\begin{equation}
K=\sum_{i=1}^r\lambda_{n_i},
\end{equation}
with $1\! \le\!  n_1\! <n_2\! <\! \cdots\! <n_r\! \le\!  N-1$. Such an eigenvalue is associated with the symmetry class $\left[N-r,1,\ldots,1\right]$, with a multiplicity of  ${N-1 \choose r}$. Most importantly, the spectral gap of  $\sigma\left(V^{(1,1,\ldots,1)}\right)$ is given by 
\begin{equation}
\label{specgap}
K_2=2d-K_{N!-1}=\lambda_2,
\end{equation}
 where $K_{N!-1}$ is the second largest eigenvalue of $\sigma\left(V^{(1,1,\ldots,1)}\right)$. Eq.~\eqref{specgap} means for instance that the spectral gap of the Cayley graph in Fig.~\ref{phedron} is equal to the one of the path graph in Fig.~\ref{path}.  This quantity can be related to the energy gap of the system through Eq.~\eqref{perten}. In the context of adiabatic quantum computing, the minimum value $\epsilon_{\mathrm{min}}$ of the energy gap along an adiabatic process is related to the minimum runtime $T$ of the algorithm by $T=O(1/\epsilon_{\mathrm{min}}^2)$ \cite{Albash2018}. It is therefore crucial to ensure that this gap remains sufficiently large along the adiabatic path, which is a priori an exponentially hard problem. Here, we have shown that it is sufficient to compute the lowest non-zero eigenvalue of the $N\times N$ matrix $V^{(N-1,N)}$ defined in Eq.~\eqref{matpath}, instead of  the whole $V^{(1,1,\ldots,1)}$ matrix, of size $N!\times N!$ \cite{notegap}. Thus, by providing an efficient way to compute the energy gap for many configurations of a many-body setting, our method constitutes a huge practical advantage.
 
% a good characterization of the energy gap is essential, since it constrains the runtime of the algorithm, but is also a hard problem for the strongly correlated systems that are typically involved \cite{Albash2018}. Here, our method constitutes a huge speed up, since it is sufficient to compute the lowest non-zero eigenvalue of the $N\times N$ matrix $V^{(N-1,N)}$ defined in Eq.~\eqref{matpath}, instead of  the whole $V^{(1,1,\ldots,1)}$ matrix, of size $N!\times N!$. For the former, \cite{COAKLEY2013379} offers an $O(N\log N)$ algorithm. 
 
\begin{figure}
	\includegraphics[width=1\linewidth]{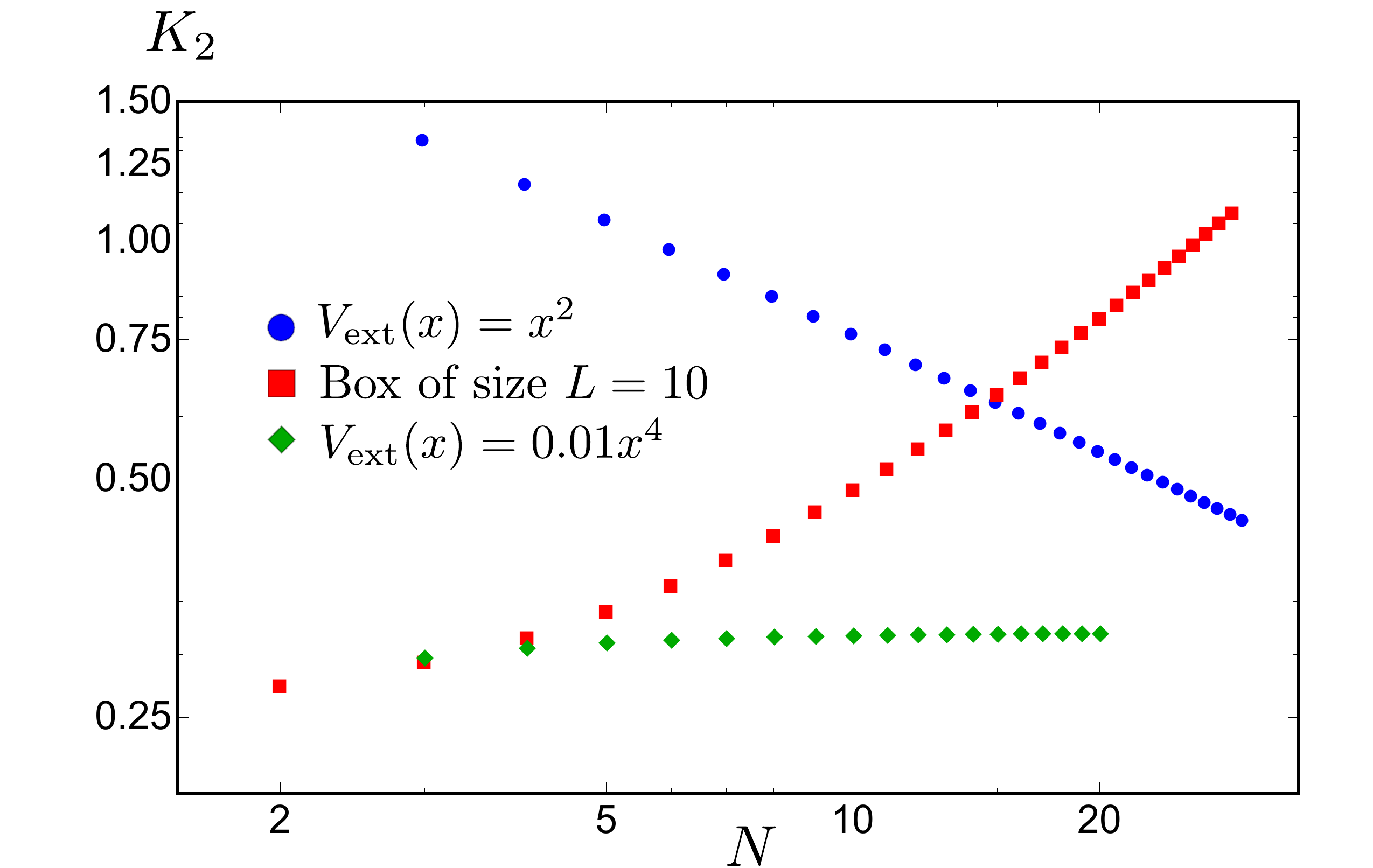}
	\caption{\label{gap}(color online). Spectral gaps $K_2$ as a function of $N=\kappa$, in log-log scale, in the cases of a box trap (red squares), harmonic (blue circles), and quartic (green diamonds) potentials. Values for the $\alpha_k$ coefficients (Eq.~\eqref{alphavolo}) come respectively from \cite{Volosniev2017}, \cite{Loft2016a} and the CONAN software \cite{Loft2016b}. Our approach allows us to drastically reduce the complexity of the problem: For $N=30$,  $\dim V^{(1,1,\ldots,1)}=30!\approx 10^{33}$ while $\dim V^{(29,1)}=30$.}
\end{figure}

 In the case of a box potential of size $L$, we can use the exact expression of $\alpha_1=\cdots=\alpha_{N-1}$ \cite{Volosniev2017} and obtain:
 \begin{equation}
 \label{alphabox}
  	K_2^{\mathrm{box}}=\frac{\pi^2N(N+1)(2N+1)}{3L^3}\left(1-\cos\frac{\pi}{N}\right).
 \end{equation}
Quite surprisingly, we note that in this case the spectral gap behaves as $K_2^{\mathrm{box}} \sim \pi^4 N/(3L^3)$ in the large $N$ limit, and is thus an {\em increasing} function of the number of particles. As displayed in Fig.~\ref{gap}, this scaling behaviour is in net contrast with the cases of a harmonic potential, where $K_2$ is a decreasing function of $N$, or a quartic potential, where $K_2$ is almost constant. We observe that the spectral gap is larger for more confining potentials, i.e. potentials $V_{\mathrm{ext}}(x)$ that grow faster when $x$ approaches infinity.

\section{Discussion}

Our results suggest that strongly confined 1D $SU(\kappa=N)$ systems may become promising candidates for quantum adiabatic computing, allowing: (i) Complex encoding due to the large number of spins; (ii) An adiabatic tuning of the exchange constants of the effective spin chain through the  external potential; (iii) A well-controlled energy gap that can be computed efficiently with our method. This paves the way to further theoretical and experimental investigations that may eventually lead to technological applications \cite{note3sm}.

There are also many theoretical perspectives to this study. For instance, one may combine this framework with random matrix theory in order to study the subtle interplay between disorder and interactions, by considering the graphs we have introduced in this work with a random choice of weights $\alpha_k$ accounting for the presence of a random external potential \cite{Abanin2019}.  Furthermore, noting that the case when the external potential is homogeneous is integrable \cite{Guan2013}, our connection suggests that these graphs play an important role in quantum integrable systems, and, conversely, that one could apply results derived from this theory to the study of Cayley graphs. More generally, this hitherto unobserved link between different communities, ranging from cold atom and condensed matter physics to pure mathematics (for the graph-theoretical and probabilistic interpretations) and even quantum technologies, should stimulate fruitful collaborations.
\begin{acknowledgments}
	We thank D. Zhang and T. Nicholson for useful discussions. This work has been partially funded by the Singapore NRF Grant numbers NRF-NRFI2017-04 (WBS No. R-144-000-378-281) and NRF-NRFF2018-02.
\end{acknowledgments}

\appendix

\section{Basic notions of graph theory}

Here we recall a few basic definitions of graph theory \cite{Bondy2008}. A \textit{graph} 
is a pair $\mathcal{G}=(\mathcal{V},\mathcal{E})$ where $\mathcal{V}$ is a set of \textit{vertices} and $\mathcal{E}$ is a set of \textit{edges} that are elements of $\mathcal{V}\times\mathcal{V}$. We say that $\mathcal{G}$ is \textit{undirected} when $(i,j)\in\mathcal{E}$ if and only if $(j,i)\in\mathcal{E}$.
The \textit{degree} $\mathrm{deg}(v)$ of a vertex $v$ is the number of vertices that are connected by an edge, or \textit{adjacent}, to $v$.
A \textit{weighted graph} is a graph $\mathcal{G}$ where each edge  $e\in\mathcal{E}$ is associated with a positive real number $w_e$ --- the unweighted case corresponding to $w_e=1$ for all $e$. In this case $\mathrm{deg}(v)$ is the sum of the weights of the edges between $v$ and its adjacent vertices. A graph is said to be \textit{regular} when all his vertices have the same degree. It is said to be \textit{bipartite} when $\mathcal{V}$ can be divided into two disjoint sets $\mathcal{V}_1$ and $\mathcal{V}_2$ such that each edge connects an element of $\mathcal{V}_1$ with an element of $\mathcal{V}_2$. If all the vertices of $\mathcal{V}_1$ (resp. $\mathcal{V}_2$) have the same degree $x$ (resp. $y$), the graph is said to be \textit{$(x,y)$-biregular}. A graph $\mathcal{G}_1=(\mathcal{V}_1,\mathcal{E}_1)$ is a \textit{covering graph} of $\mathcal{G}_2=(\mathcal{V}_2,\mathcal{E}_2)$ if there is a surjective map $f:\mathcal{V}_1\to\mathcal{V}_2$ such that, for each $v\in\mathcal{V}_1$, the restriction of $f$ to a neighbourhood of $v$ is a bijection onto a neighbourhood of $f(v)$.

Given a graph $\mathcal{G}=(\mathcal{V},\mathcal{E})$, one can canonically associate a set of $|\mathcal{V}|\times|\mathcal{V}|$ matrices, $|\mathcal{V}|$ being the number of vertices, whose spectra can be related to the structural properties of the graph \cite{Brouwer2012}. The \textit{adjacency matrix} $A_{\mathcal{G}}$ of $\mathcal{G}$ is such that entry $A_{ij}$ is  equal to $1$ (or $w_{(i,j)}$ if the graph is weighted) if vertex ``$i$'' is adjacent to vertex ``$j$'' and $0$ otherwise. The \textit{degree matrix} $D_{\mathcal{G}}$ is simply the diagonal matrix whose diagonal entries are given by the degree of the corresponding vertex. The \textit{Laplacian matrix} is then defined by
\begin{equation}
\Delta_{\mathcal{G}}=D_{\mathcal{G}}-A_{\mathcal{G}}.
\end{equation}
This matrix can be seen as  a discrete version of the (negative) continuous Laplacian operator, hence the name. It can also be interpreted as the generator of a random walk on $\mathcal{V}$ \cite{Lovász93randomwalks}.

Let us now enunciate some basic properties of the Laplacian matrix. It is easy to see that, for an undirected graph, $\mathcal{G}$, $\Delta_{\mathcal{G}}$ is symmetric and therefore it can be diagonalized in an orthonormal basis and has a real  spectrum $\sigma\left(\Delta_{\mathcal{G}}\right)\subset\mathbb{R}$. Moreover, the fact that all the diagonal entries of $\Delta_{\mathcal{G}}$ are positive and that each diagonal entry is equal to the sum of the absolute values of the non-diagonal entries in that row imply that $\sigma\left(\Delta_{\mathcal{G}}\right)\subset\mathbb{R}_+$. Furthermore, since every row sum and column sum is zero,
the vector $u=(1,1,\ldots,1)\in\mathbb{R}^{|\mathcal{V}|}$ always satisfies $\Delta_{\mathcal{G}}u=0$, and thus   $0\in\sigma\left(\Delta_{\mathcal{G}}\right)$. Noting that if $\mathcal{G}$ has $m$ connected components \cite{note1}, $\Delta_{\mathcal{G}}$ is a block-diagonal matrix with $m$ blocks, we see that the multiplicity of $0$ is equal to $m$. In particular, for a connected graph, $\sigma\left(\Delta_{\mathcal{G}}\right)$ can be written:
\begin{equation}
0=\lambda_1<\lambda_2\le\cdots\le\lambda_{|\mathcal{V}|}.
\end{equation} 

Many bounds have been obtained on these eigenvalues, and in particular on $\lambda_{|\mathcal{V}|}$ and on the \textit{spectral gap} $\lambda_2$, which can be related to geometrical properties of $\mathcal{G}$ (see e.g. \cite{Chung1996} or \cite{Godsil2001}). For example, one has \cite{Mohar1991}
\begin{equation}
\label{biregmax}
\lambda_{|\mathcal{V}|}\le\max\{\mathrm{deg}(u)+\mathrm{deg}(v);(u,v)\in\mathcal{E}\},
\end{equation}
with equality if and only if $\mathcal{G}$ is biregular (in the case of a connected graph).

Finally, one can get the spectrum of graphs obtained by operations on other graphs with a known spectrum \cite{Mohar1991}. For example, given two graphs $\mathcal{G}_1=(\mathcal{V}_1,\mathcal{E}_1)$ and $\mathcal{G}_2=(\mathcal{V}_2,\mathcal{E}_2)$, we can define the \textit{Cartesian product}  $\mathcal{G}_1\times \mathcal{G}_2$ with vertices in $\mathcal{V}_1\times \mathcal{V}_2$ and such that there is an edge between $(a,b)$ and $(a',b')$ if either $(a,a')\in\mathcal{E}_1$ or $(b,b')\in\mathcal{E}_2$. Then, writing $\sigma\left(\Delta_{\mathcal{G}_1}\right)=(\lambda_i)_{1\le i \le k}$ and $\sigma\left(\Delta_{\mathcal{G}_2}\right)=(\mu_i)_{1\le i \le l}$, one has
\begin{equation}
\label{cartprod}
\sigma\left(\Delta_{\mathcal{G}_1\times\mathcal{G}_2}\right)=(\lambda_i+\mu_j)_{1\le i \le k,1\le j \le l}.
\end{equation}
This formula is crucial in Bacher's proof for the unweighted version of Eq.~(11) of the main text \cite{Bacher1994}.

\section{Representation theory of $\mathfrak{S}_N$}
\label{aprep}

In what follows we describe the construction of the set $\mathrm{IR}$ of irreducible representation (irreps) of the symmetric group $\mathfrak{S}_N$ (see e.g. \cite{James1984}).

The set $\mathrm{IR}$ is in bijection with the \textit{conjugacy classes} of $\mathfrak{S}_N$, which are characterized by a given structure for their decomposition in disjoint cyclic permutations. Thus, there is a one-to-one correspondence between $\mathrm{IR}$ and the set of partitions of $N$. A convenient way of representing a partition \cite{note2a} $\mu=\left[\mu_1,\ldots,\mu_r\right]$ of $N$ is through a \textit{Young diagram}, a left-justified set of boxes with $r$ rows, where each row
$i\in\{1,\ldots,r\}$ contains $\mu_i$ boxes. For example, an irrep of $\mathfrak{S}_8$ is characterized by the partition $\left[4,3,1\right]$, or equivalently by the following Young diagram:
\begin{equation}
Y_{\left[4,3,1\right]}\equiv\yng(4,3,1).
\end{equation}
$\left[4,3,1\right]$ is called the \textit{shape} of $Y_{\left[4,3,1\right]}$, and we denote the corresponding (class of) irrep(s) by $\rho_{\left[4,3,1\right]}$. The \textit{conjugate} of a Young diagram of shape $\mu=\left[\mu_1,\ldots,\mu_r\right]$ is the diagram with columns of lengths $\mu_1,\ldots,\mu_r$. E.g. the conjugate of the previous example is
\begin{equation}
Y_{\left[3,2,2,1\right]}\equiv\yng(3,2,2,1).
\end{equation}

The set of Young diagrams is partially ordered by the so-called \textit{dominance order} $\trianglerighteq$, such that $\mu\equiv\left[\mu_1,\ldots,\mu_r\right]\trianglerighteq\nu\equiv\left[\nu_1,\ldots,\nu_r\right]$ (where the last terms of one of the partitions may be equal to zero) if and only if
\begin{equation}
\mu_1+\cdots+\mu_k\ge\nu_1+\cdots+\nu_k\quad\text{for all }1\le k\le r.
\end{equation}
For example, one has
\begin{equation}
\yng(4,3,1)\quad\trianglerighteq\quad\yng(3,2,1,1,1)~.
\end{equation}
Intuitively, it means that one can go from the left diagram to the right one by moving a certain number of boxes from upper rows to lower rows. Note that it is not a total order when $N>5$. For instance, it is not possible to compare
\begin{equation}
\yng(4,1,1)\quad\text{and}\quad\yng(3,3)~.
\end{equation}

\vspace{0.5cm}

Let us be a little more precise on how an irrep can be constructed from a Young diagram.  A \textit{Young tableau} is a Young diagram whose boxes are labelled by integers. Two tableaux of same shape $\mu$ are said to be \textit{row-equivalent} if they are equal up to permutations of the rows. For example,
\begin{equation}
\young(8451,372,6)\quad\text{and}\quad\young(1854,732,6)
\end{equation}
are row-equivalent. This defines equivalence classes on the set of $\mu$-tableaux that are called \textit{tabloids}. For the previous example, it is represented as 
\ytableausetup{tabloids,centertableaux}
\begin{equation}
\label{extabloid}
\ytableaushort{1458,237,6}~.
\end{equation}
Note that each permutation of a fermionic mixture, or snippet, can be regarded as a tabloid with rows of lengths $N_1,\ldots,N_{\kappa}$ and with entries of row $i$ corresponding to the positions of the particle of type $i$. For example, if we consider an $8$-particle mixture with $4$ particles of type $a$, $3$ particles of type $b$ and $1$ of type $c$, the tabloid given in Eq.~\eqref{extabloid} corresponds to the following  snippet:
\begin{equation}
\label{extabloid2}
\ytableaushort{1458,237,6}\quad\leftrightarrow\quad abbaacba.
\end{equation}

We now define the \textit{permutation module} as the $\mathbb{C}$-vector space $M^{\mu}$ whose basis is indexed by the set of $\mu$-tabloids. Note that its dimension is $D_\mu$, as defined in the main text. Note also that $M^{\left[1,1,\ldots,1\right]}$ is in bijection with $\mathbb{C}\left[\mathfrak{S}_N\right]$, the \textit{group algebra} of $\mathfrak{S}_N$, whose basis $(e_P)_{P\in \mathfrak{S}_N}$ is indexed by the elements of $\mathfrak{S}_N$ and such that $e_P\cdot e_Q =e_{PQ}$. Associated with $\mathbb{C}[\mathfrak{S}_N]$ is the \textit{(left) regular representation} $\rho$ of $\mathfrak{S}_N$: writing a vector in $\mathbb{C}[\mathfrak{S}_N]$ as $u=\sum_{Q\in \mathfrak{S}_N}u_Qe_Q$ with $u_Q\in \mathbb{C}$, and given $P\in \mathfrak{S}_N$, the linear map $\rho(P)$ is defined by
\begin{equation}
\rho(P)\cdot u=e_{P}u=\sum_{Q\in \mathfrak{S}_N}u_Qe_{PQ}=\sum_{Q\in \mathfrak{S}_N}u_{P^{-1}Q}e_{Q},
\end{equation}
which can be linearly extended on $\mathbb{C}[\mathfrak{S}_N]$ by writing $\rho\left(\sum_{Q\in \mathfrak{S}_N}u_Qe_Q\right)\equiv\sum_{Q\in \mathfrak{S}_N}u_Q\rho(Q)$. Similarly, one can identify any permutation module $M^{\mu}$ with a representation of $\mathfrak{S}_N$ by considering the natural action of $\mathfrak{S}_N$ on the vector space $M^{\mu}$.

Then, for a given $\mu$-tableau $T$, one can associate the following element of $M^{\mu}$:
\begin{equation}
\label{etspecht}
E_T=\sum_{P\in C_T}\epsilon(P)\{P(T)\},
\end{equation}
where $C_T$ is the subgroup of permutations preserving all columns of $T$, $\epsilon(P)$ is the sign of the permutation $P$, and $\{T\}$ is the tabloid corresponding to $T$. $E_T$ is called a \textit{polytabloid}. Then, the \textit{Specht module} $S_\mu$ is defined as the subspace of $M_{\mu}$ generated by the elements $E_T$ when $T$ runs through all the $\mu$-tableaux. It can be shown that the basis of $S^\mu$ is given by the elements $E_T$ when $T$ runs though all the \textit{standard} Young tableaux of shape $\mu$, that is the tableaux whose entries are increasing from left to right along the rows and up to down along the columns. In particular, its dimension is given by the number of standard Young tableaux, which can be easily obtained from the so-called \textit{hook length formula}. Furthermore, the so-called \textit{Young's rule} states that the permutation module $M^\mu$ can be decomposed in the following way:
\begin{equation}
\label{youngsrule}
M^\mu\cong\bigoplus_{\mu'\trianglerighteq\mu}k_{\mu'\mu}S^{\mu'},
\end{equation}
where $k_{\mu'\mu}$ are positive integers.

Here again, $S^\mu$ can be seen as a representations of $\mathfrak{S}_N$. Then, it can be shown that the set of all the Specht modules $S^\mu$ is, in fact, $\mathrm{IR}$. Note that when taking  $\mu=\left[1,1,\ldots,1\right]$, Eq.~\eqref{youngsrule} implies in particular that the regular representation is a sum of all the irreps. In this case, the $k_{\mu'\left[1,1,\ldots,1\right]}$ numbers are given by the dimensions of the $S^{\mu'}=\rho_{\mu'}$ irreps. Intuitively, keeping in mind Eq.~\eqref{etspecht}, an irrep $\rho_{\mu}$ can be seen as symmetrizing the rows and anti-symmetrizing the columns of the $\mu$-tableaux.

\section{Generalized Lieb-Mattis theorem --- A detailed proof of $K_{max}^{[\mu]} = K_{max}^{\mu}$}

\label{aplm}

Denoting by $K_{max}^{[\mu]}$ the maximum eigenvalue of $V^{(1,1,\ldots,1)}$ corresponding to the symmetry class $[\mu]$, and by $K_{max}^{\mu}$ the maximum eigenvalue of $V^{\mu}$, we have shown in the main text that $K_{max}^{[\mu]}\in\sigma\left(V^{\mu}\right)$, and then we claimed that in fact 
\begin{equation}
\label{mumu}
K_{max}^{[\mu]} = K_{max}^{\mu}.
\end{equation}

In order to prove this assertion, let us first observe that the graphs $X(\mathfrak{S}_{\mu}\subset \mathfrak{S}_N,S_c)$ are bipartite. This is immediately clear for the Cayley graph $X(\mathfrak{S}_N,S_c)$, where the vertices associated with   $P\in\mathfrak{S}_N$ are separated between the ones corresponding to $\epsilon(P)=1$ and the ones corresponding to $\epsilon(P)=-1$. In the general case of a Schreier graph, the vertices are the left cosets $P \mathfrak{S}_{\mu}$, or equivalently the set of tabloids. With each $P \mathfrak{S}_{\mu}$, we can associate the representative $\overline{P}\in P \mathfrak{S}_{\mu}$ such that the indices corresponding to particles belonging to a same component are displayed in increasing order. For example, the representative of the tabloid displayed in Eq.~\eqref{extabloid2} corresponds to
\begin{equation}
abbaacba \quad\leftrightarrow\quad 15623874,
\end{equation}
with particles $1,2,3,4$ of type $a$, $5,6,7$ of type $b$, and $8$ of type $c$.
Defining the sign of $P\mathfrak{S}_{\mu}$ by $\mathrm{sign}(P \mathfrak{S}_{\mu})\equiv\epsilon(\overline{P})$, we see that the vertices of $X(\mathfrak{S}_{\mu}\subset \mathfrak{S}_N,S_c)$ are also separated between the positive and the negative ones.

Given a mixture $\mu$, we consider an eigenvector $a$ of $V^{\mu}$ with eigenvalue $K_{max}^{\mu}$. Then, we can define a vector $\tilde{a}$ such that, for every component $i\in\{1,\ldots,D_{\mu}\}$ corresponding to some coset $P_i \mathfrak{S}_{\mu}$:
\begin{equation}
\tilde{a}_i=\mathrm{sign}(P_i \mathfrak{S}_{\mu})|a_i|.
\end{equation}
Seeing $a$ as an element of the permutation module $M^{\mu}$, we deduce from the definition of $\mathrm{sign}(P \mathfrak{S}_{\mu})$ that $\tilde{a}$ is an element of the Specht module $S^\mu$, i.e. that the corresponding state belongs to the symmetry class of type $\left[\mu\right]$. Moreover, since $V^{\mu}$ is a real symmetric matrix, we can write $\tilde{a}$ as a sum of orthogonal eigenvectors $\tilde{a}^j$ with eigenvalues $K^j$. Therefore, we have
\begin{equation}
\left\|\tilde{a}\right\|^2=\sum_j\left\|\tilde{a}^j\right\|^2=\left\|a\right\|^2
\end{equation}
and
\begin{equation}
\left\|V^{\mu}a\right\|^2=(K_{max}^{\mu})^2\left\|a\right\|^2\le\sum_j (K^j)^2\left\|\tilde{a}^j\right\|^2=\left\|V^{\mu}\tilde{a}\right\|^2.
\end{equation}
Since $K_{max}^{\mu}$ is the maximum eigenvalue of $V^{\mu}$, we deduce that $\tilde{a}$ belongs to the eigenspace with eigenvalue $K_{max}^{\mu}$ and we obtain Eq.~\eqref{mumu}.

In fact, we can prove that, \textit{in generic cases}, the only vectors with maximum eigenvalue $K_{max}^{\mu}$ belong to the symmetry class $\left[\mu\right]$. This can be seen as a consequence of a general theorem \cite{Poignard2018}, which states that, for a given weighted graph, the spectrum of its Laplacian matrix is simple for generic choices of weights \cite{noteleb}. In our situation, this means that a small random perturbation of the external potential $V_{ext}(x)$ will lead to a spectrum $\sigma\left(V^{\mu}\right)$ that contains no accidental degeneracies between different symmetry classes --- although it usually does contain degeneracies associated with $k_{\mu'\mu}$, which denote the number  of time the irrep $S^{\mu'}$  appears in the decomposition of the permutation module $M^{\mu}$ according to Young's rule (cf Eq.~\eqref{youngsrule}).

\vspace{0.5cm}

Using the decomposition of the spectrum according to the irreps of $\mathfrak{S}_N$ as described in the main text, Eq.~\eqref{mumu} implies that
\begin{equation}
\label{klieba}
\mu\trianglerighteq\mu'\quad\implies\quad K_{max}^{[\mu]}\le K_{max}^{[\mu']}.
\end{equation}
Writing $E_0^{[\mu]}(1/g_{\mathrm{1d}})=E_A-K_{max}^{[\mu]}/g_{\mathrm{1d}}$ the energy of the corresponding ground state with symmetry $\left[\mu\right]$ in the strongly repulsive limit, we thus have:
\begin{equation}
\mu\trianglerighteq\mu'\quad\implies\quad E_0^{[\mu]}(1/g_{\mathrm{1d}})\ge E_0^{[\mu']}(1/g_{\mathrm{1d}})\quad(g_{\mathrm{1d}}\gg 1).
\end{equation}
This is a generalized version of the Lieb-Mattis theorem, which has important consequences in the theory of magnetism  \cite{LiebMattisPR}. Intuitively, the ground-state wavefunction ``wants'' to be as symmetric as possible. For spin-$1/2$ particles ($\kappa=2$ case), this means that the total spin of the system is minimized, and thus that the ground-state is unmagnetized. Eq.~\eqref{klieba} has been checked for several few-body systems \cite{Decamp2016a,Decamp2016b}. Our approach provides a rigorous and general proof.

Note that Eq.~\eqref{mumu} provides a way to compare ground-state energies $E_0^{[\mu]}(1/g_{\mathrm{1d}})$, $E_0^{[\mu']}(1/g_{\mathrm{1d}})$ when $\mu$ and $\mu'$ are not comparable according to the dominance order. It is indeed sufficient to compare the spectral radii $K_{max}^{\mu}$ and $K_{max}^{\mu'}$ of the real symmetric matrices $V^\mu$ and $V^{\mu'}$ (resp.), which can be obtained effectively, e.g. using the Lanczos algorithm \cite{Lanczos:1950zz}.

\section{Proposed experimental implementation}

\label{apexp}

The model of strongly-correlated one-dimensional spinor gases can be realized in the laboratory by working with group-II atoms like strontium and ytterbium. These atoms in their singlet ground state only have nuclear spin components, therefore they exhibit $SU(\kappa)$ symmetry \cite{Gorshkov2010, Ye2014, Foelling2014, Pagano2014}. To prepare a specific number of atoms $N$, one can use optical tweezers to prepare one atom per tweezer in the collisional blockade regime \cite{Grangier2001}, and use acousto-optical modulators \cite{Lukin2016}, spatial light modulators \cite{Browaeys2016}, or microlens arrays \cite{Birkl2019} to scale up to a desired number of tweezers. Alternatively, one can use a quantum gas microscope to determine the number of atoms loaded into isolated chains \cite{Greiner2009, Bloch2010}. To obtain a specific superposition of spins, one can coherently address different spin states using clock transitions with long-lived coherence times. Having prepared the atom number and quantum states, these atoms can then be loaded into an overall 1D potential while the tweezer or lattice potentials separating the individual atoms are turned off. A harmonic 1D potential can be created by interfering red-detuned lattice beams in two dimensions. A confinement-induced resonance, where the transverse confinement length is tuned to match the scattering length, can be used to reach the strongly repulsive regime $g_{\mathrm{1d}} \rightarrow \infty$ \cite{Jochim2012}. To vary $\alpha_k$, one can change the external 1D potential. For example, instead of a harmonic potential, a box potential can be created with the help of steep repulsive walls formed from blue-detuned light shaped by spatial light modulators \cite{Hadzibabic2013} or other diffractive optics \cite{Zwierlein2017}. 

The contact $K$ can be determined from a range of experimental methods, including Bragg spectroscopy \cite{Vale2010, Vale2011}, RF spectroscopy \cite{Stewart2010,Cornell2012, Jin2012, Zwierlein2019}, RF Ramsey interferometry \cite{Fletcher2017}, and photoassociation experiments \cite{Castin2009, Hulet2005}. Of these, measuring the molecular fraction from photoassociation is likely to be one of the most sensitive methods \cite{Ni2018} in the limit of one atom per spin component, where $N = \kappa$. Beyond the ground state, the contact for the first excited state $K_{N!-1}$ will become accessible as the temperature of the ultracold atoms increases. In particular, our graph theory analysis predicts the difference between the maximum contact $K_{max}$ and the second largest contact $K_{N!-1}$ where $N = \kappa$ to be the same as $K_2$ for the case of $N-1$ particles in spin down and one particle in spin up. Such symmetry allows us to gain insight on spectral gaps in the latter case, especially for highly excited many-body states that are otherwise experimentally challenging to access.


\begin{thebibliography}{100}
\bibitem{notespinchain} Note that $V^{\nu}$ can be written as a function of the generators of the $SU(\kappa)$ algebra (i.e. the Pauli matrices when $\kappa=2$), which allows a mapping to a spin chain with coupling constants given by $J_k=\alpha_k/g_{\mathrm{1d}}$ \cite{Deuretzbacher2014}.
\bibitem{note1} Note that since $S$ generates $G$, $X(G,S)$ is always connected. If $S$ is symmetric, that is $s\in S\Leftrightarrow s^{-1}\in S$, then $X(G,S)$ is undirected and $d$-regular, with $d\equiv\sum_{s\in S}w_s$.
\bibitem{note2sm} See Appendix \ref{aplm} for our detailed proof.
\bibitem{note2a} We chose a slightly different notation for the partitions here in order to avoid confusion with partitions $(N_1,\ldots,N_{\kappa})$ characterizing  the choice of the mixture.
\bibitem{noteleb} More precisely, the set of weights of a graph $\mathcal{G}$ such that $\sigma\left(\Delta_{\mathcal{G}}\right)$ is simple has full Lebesgue measure.
\end{thebibliography}
\end{document}